\pdfoutput=1
\RequirePackage{ifpdf}
\ifpdf 
\documentclass[pdftex]{sigma}
\else
\documentclass{sigma}
\fi

\begin{document}

\allowdisplaybreaks

\newcommand{\arXivNumber}{1906.04939}

\renewcommand{\thefootnote}{}

\renewcommand{\PaperNumber}{020}

\FirstPageHeading

\ShortArticleName{Geometric Approach to Quantum Theory}

\ArticleName{Geometric Approach to Quantum Theory\footnote{This paper is a~contribution to the Special Issue on Algebra, Topology, and Dynamics in Interaction in honor of Dmitry Fuchs. The full collection is available at \href{https://www.emis.de/journals/SIGMA/Fuchs.html}{https://www.emis.de/journals/SIGMA/Fuchs.html}}}

\Author{Albert SCHWARZ}

\AuthorNameForHeading{A.~Schwarz}

\Address{Department of Mathematics, UC Davis, Davis, CA 95616, USA}
\Email{\href{mailto:schwarz@math.ucdavis.edu}{schwarz@math.ucdavis.edu}}
\URLaddress{\url{http://www. math.ucdavis.edu/~schwarz}}

\ArticleDates{Received February 29, 2020, in final form March 25, 2020; Published online April 01, 2020}

\Abstract{We formulate quantum theory taking as a starting point the cone of states.}

\Keywords{state; cone; quantum}

\Classification{81P05}

\begin{flushright}
\begin{minipage}{70mm}
\emph{To Dmitry Fuchs, a brilliant mathematician and a wonderful human being}
\end{minipage}
\end{flushright}

\renewcommand{\thefootnote}{\arabic{footnote}}
\setcounter{footnote}{0}

\section{Introduction}
The starting point of algebraic quantum theory is a unital associative algebra $\mathcal A$ with antilinear involution $A\to A^*$. The observables are self-adjoint elements of the algebra (fixed points of the involution). States are defined as positive linear functionals on~$\mathcal A$. (One says that a linear functional $\omega $ is positive if $\omega (A^*A)\geq 0$ for~$A\in \mathcal A$. It is normalized if $\omega (1)=1$.) States constitute a convex cone (denoted by $\mathcal C$) in the space~$\mathcal L$ of linear functionals obeying $\omega (A^*)=\overline{\omega (A)}$.

We are taking as a starting point a convex cone $\mathcal C$ in a normed space (or, more generally, in topological vector space)~$\mathcal L$. Elements of $\mathcal C$ are called states. We fix also a linear functional $e$ on~$\mathcal L$ that allows us to define a normalized state as a state $\omega$ obeying $e(\omega)=1$. We show how one can define observables and probabilities using these data.

It was noticed many years ago that the associative algebra $\mathcal A$ does not have any direct physical meaning because a product of self-adjoint elements is not necessarily self-adjoint. The remark that the anticommutator $AB+BA$ of two self-adjoint elements is self-adjoint led to the notion of Jordan algebra \cite{HO, JVW}. Jordan algebras and Jordan superalgebras were studied in numerous papers, the first classification result was proven in~\cite{JVW}, classification of Jordan superalgebras studied by Kac and Zelmanov~\cite{KAC,KMZ}.

One can define the set of non-negative elements of Jordan algebra as the set of all elements that can be represented as squares. Under some conditions one can prove that this set is a~convex cone. Then we can consider the corresponding quantum theory.

One can construct quantum theory starting with any convex cone. It seems, however, that the most interesting theories correspond to homogeneous cones (an open cone is called homogeneous if the group of automorphisms acts transitively). Homogeneous cones and corresponding homogeneous complex domains were studied in numerous papers. In finite-dimensional case they were classified by E.~Vinberg \cite{VV,VVV,VGP,XU}.
In particular, self-dual cones correspond to Euclidean Jordan algebras.

One can hope that these beautiful results will lead to new interesting examples of quantum theories.

The probabilities in conventional formulation of quantum mechanics can be derived from decoherence; see, for example~\cite{SCT}. In this geometric approach we introduce the probabilities axiomatically, however, one can apply the considerations of \cite {SCT} to derive them from first principles (at least for integrals of motion). (In our approach the prescription for calculation of probabilities was first derived from decoherence and later axiomatized.)

In the follow up paper we introduce the notion of particle and develop scattering theory in geometric approach generalizing the results of~\cite{SC}.  We apply this generalization to formulate the scattering theory in terms of Jordan algebras. One can hope that working in terms of Jordan algebras one can get a~deeper understanding of superstring. (A partially successful attempt to relate the exceptional Jordan algebra to superstring was made in~\cite{FJ}.)

\section{Quantum theory: states and observables}

Let $\mathcal C$ be a cone in a normed complex vector space $\mathcal L$. (It is easy to generalize our constructions to the case when $\mathcal L$ is a topological vector space.) We define a cone as a convex subset such that for every $\lambda\geq 0$ and $x\in \mathcal C$ we have
$\lambda x\in \mathcal C$. (In other words, the cones we consider are always convex cones.) Equivalently one can say that a cone is a subset of $\mathcal L$ such that $\lambda x+\mu y\in \mathcal C$ for $x,y\in {\mathcal C}$, $\lambda, \mu\geq 0$. We assume that the cone $\mathcal C$ is a~closed set and that it can be considered as a~closure of open cone.

We fix a linear functional $e$ on $\mathcal L$ (normalizing functional). We say that a state $\omega\in \mathcal C$ obeying $e(\omega)=1$ is normalized. The set of normalized states will be denoted by ${\mathcal C}_0$; {\it we assume that this set is bounded}.

Automorphisms of the cone~$\mathcal C$ are defined as invertible bicontinuous linear operators on $\mathcal L$ inducing a bijective map of the cone~$\mathcal C$ onto itself. The group of automorphisms will be denoted by $\operatorname{Aut}$. We will be interested first of all in the subgroup of $\operatorname{Aut}$ consisting of automorphisms that preserve the normalizing functional~$e$; this subgroup will be denoted by~$\mathcal U$. It consists of automorphisms of ${\mathcal C}_0$.

The evolution is specified by a family $\sigma (t) \in \mathcal U$ (a family of automorphisms of ${\mathcal C}_0$). The Hamiltonian $H(t) $ should be regarded as an infinitesimal automorphism:
\[
{\rm i} \frac {{\rm d}\sigma(t)}{{\rm d}t}=H(t)\sigma (t).
\]
If $H(t)=H$ does not depend on time we write $\sigma(t)={\rm e}^{-{\rm i}Ht}$.

More generally a physical quantity $A$ specifies an infinitesimal element of $\mathcal U$ (i.e., it generates a one-parameter group $\alpha(t)$ of automorphisms of $\mathcal C$ preserving~$e$.)

We consider a physical quantity as a pair $(A, a)$ where $ A$ is an infinitesimal automorphism
and~$a$ is a linear functional that is invariant with respect to~$\alpha(t)$.

In standard formulation a self-adjoint operator $\hat A$ on Hilbert space $\mathcal H$ determines an operator~$A$ on density matrices and linear functional~$a$:
\begin{gather*}
A(K)=\hat AK-K\hat A,\\
a(K)=\operatorname{Tr} \hat A K.
\end{gather*}

The group $\mathcal U$ can be identified with the group of unitary transformations of $\mathcal H$ acting on density matrices by the formula $K\to UKU^{-1}$. The infinitesimal elements of $\mathcal U$ correspond to self-adjoint operators in~$\mathcal H$. The family of automorphisms~$\alpha (t)$ is given by the formula
\[
\alpha (t)K= {\rm e}^{-{\rm i}At}K={\rm e}^{-{\rm i}\hat At}K{\rm e}^{{\rm i}\hat At}.
\]

\section{Probabilities}
For a physical quantity $(A,a)$ we introduce an operator
\[
 Q_A=\lim_{\epsilon\to 0}\int \frac {\sqrt {\epsilon}}{\sqrt{2\pi}} {\rm e}^{-\epsilon t^2} \alpha(t){\rm d}t
 \]
mapping the set of normalized states ${\mathcal C}_0$ into the set of normalized states obeying
$Ax=0$ (into the set of stationary states if $A=H$).

To calculate the probabilities of the physical quantity $(A,a)$ in a normalized state $x$
we should calculate $Q_A x$ and represent it as a mixture of extreme states $s_k$ in
$Q_A{\mathcal C}_0$ (of {\it pure states}):
\[
Q_Ax=\sum p_k s_k.
\]
Then we are saying that the measurement gives us $a(s_k)$ with the probability $p_k$.

We have assumed that the set ${\mathcal C}_0$ is bounded. It follows that the automorphisms $\alpha (t)= {\rm e}^{-{\rm i}At}$ are bounded operators. This implies that all eigenvalues of $A$ are real and that there are no associate generalized eigenvectors (Jordan blocks are one-dimensional). If~$A$ has discrete spectrum we can conclude that in appropriate basis the operator~$A$ is diagonal: $A\epsilon_n=a_n\epsilon _n$, $a_n\in {\mathbb R}$. It is easy to check that $Q_A\epsilon_n=\epsilon_n$ if $a_n=0$, $ Q_A\epsilon _n=0$ if $a_n\neq 0$. (In other words~$Q_A$ is a projection on the $\operatorname{Ker} A$.)

In the standard formalism, when $A$ is a commutator with $\hat A$, the eigenvectors of~$A$ correspond to a pair of eigenvectors of~$\hat A$ and the eigenvalues of~$A$ are differences of eigenvalues of~$\hat A$. In $\hat A$-representation the basis of the space of trace class operators consists of matrices having only one non-zero entry equal to~$1$. If the spectrum of the operator~$\hat A$ is simple the operator~$Q_A$ sends all non-diagonal matrices to zero and leaves intact diagonal matrices.

\medskip

{\bf Acknowledgements.} I am indebted to V.~Kac and E.~Vinberg for very useful discussions.

\pdfbookmark[1]{References}{ref}
\LastPageEnding

\end{document}